\documentstyle[titlepage,12pt]{article}
\newcommand{\mathrmd}{{\rm d}}
\newcommand{\mathrmdt}{{\rm dt}}

\title{Computation of conserved densities \\ 
for systems of nonlinear  \\ 
differential-difference equations\thanks{Research 
supported in part by the NSF under Grant CCR-9625421.}
} 

\author{\\ \\ \\  \"{U}nal G\"{o}kta\c{s}\thanks{
E-mail: $\{$ugoktas,whereman,gerdmann$\}$@mines.edu
}, 
Willy Hereman$^2$, and Grant Erdmann$^2$ \\ \\ \\
Department of Mathematical and Computer Sciences, \\
Colorado School of Mines, Golden, CO 80401-1887, U.S.A. \\ \\ \\
Submitted to: Phys. Lett. A
}
\begin{document}
\maketitle
\vfill
\newpage
\noindent
{\bf Abstract}
\vskip 12pt
\noindent
A new method for the computation of conserved densities of nonlinear 
differential-difference equations is applied to Toda lattices and 
discretizations of the Korteweg-de Vries and nonlinear Schr\"odinger 
equations.
The algorithm, which can be implemented in computer algebra
languages such as {\it Mathematica}, can be used as an indicator of 
integrability. 


\section{Introduction}

Nonlinear differential-difference equations (DDEs) describe 
many interesting phenomena such as vibrations of particles 
in lattices, charge fluctuations in networks, Langmuir waves 
in plasmas, interactions between competing populations.
Mathematically, DDEs also occur as spatially discrete analogues 
of partial differential equations (PDEs). 
As such, lattices play a key role in numerical solvers for PDEs 
\cite{MAandBH}. 

In \cite{UGandWH,UGa,UGb,WHandWZ}, we introduced an algorithm to find 
the analytical form of polynomial conserved densities for systems of 
nonlinear evolution equations. 
We used the concept of scaling symmetries or dimensional analysis.
That inherently limits the algorithm to polynomial densities 
and fluxes of polynomial systems. 
The algorithm was implemented \cite{UGb} in {\it Mathematica}.  
Here we present its extension to semi-discrete 
polynomial systems.
We aim at deriving a set of independent conservation laws of DDEs, 
hence predicting integrability.

There are several motives to find conserved densities of DDEs explicitly.
The first few conservation laws may have a physical meaning, such as 
conserved momentum and energy. 
Additional ones may facilitate the study of both quantitative 
and qualitative properties of solutions \cite{EHandKO}.
Furthermore, the existence of a sequence of conserved densities predicts 
integrability of DDEs.
Yet, the nonexistence of conserved quantities does not preclude integrability.
Indeed, integrable DDEs could be disguised with a coordinate transformation 
so that they no longer admit conserved densities of polynomial type. 

Another compelling argument relates to the numerical solution of PDEs. 
In numerical schemes the discrete conserved quantities should 
remain constant.
In particular, the conservation of a positive definite quadratic quantity 
may prevent the occurrence of nonlinear instabilities in the numerical 
scheme. The use of conservation laws in PDE solvers has been 
discussed in \cite{FJH,RJLeV,JMS-S}.

For nonlinear DDEs several solution methods and integrability tests
are applicable. The solution methods include symmetry reduction 
\cite{DLandPW}, and an extension of the spectral transform 
method \cite{DLandOR}.   
Adaptations of the singularity confinement approach \cite{ARandBGandKT}, 
the Wahlquist-Estabrook method \cite{BD}, and the master symmetry technique 
\cite{ICandRY} allow one to test integrability of DDEs. 
In contrast, our method is completely algorithmic and can be implemented 
in computer algebra languages. 

In Section 2, the algorithm is illustrated with the Toda lattice 
\cite{MT}. 
In Section 3, we find densities for discrete analogues of the 
Korteweg-de Vries (KdV), nonlinear Schr\"odinger (NLS), and generalized 
Toda equations. These examples show subtle points of the algorithm. 
We draw conclusions in Section 4.
% \vfill
% \newpage
\section{Conserved Densities}

\subsection{Definitions}

Consider a system of DDEs that are continuous in time, and discretized 
in the (single) space variable, 
\begin{equation}\label{multisys}
{\dot{\bf u}}_n = 
{\bf F} (...,{\bf u}_{n-1}, {\bf u}_{n}, {\bf u}_{n+1},...), 
\end{equation}
where ${\bf u}_{n}$ and ${\bf F}$ are vector dynamical variables with 
any number of components. 
For simplicity of notation, the components of ${\bf u}_n $ are denoted 
by $u_n, v_n,$ etc.
We assume that ${\bf F}$ is polynomial with constant coefficients. 
If DDEs are of second or higher order in $t$, we assume that they
can be recast in the form (\ref{multisys}).

For (\ref{multisys}), we define a local {\em conservation law\/} by
\begin{equation}\label{conslaw}
{\dot{\rho}}_n = J_n - J_{n+1},
\end{equation}
where $\rho_n $ is the {\em conserved density} and $J_n$ is the 
associated {\em flux.} Both functionals are assumed to be polynomials 
in ${\bf u}_n$ and its shifts. Also, (\ref{conslaw}) is satisfied on 
solutions of (\ref{multisys}). 
Our algorithm is currently restricted to the shift-up operator $U$, where
$(I - U) J_n = J_n - J_{n+1}. $ 
Minor modifications would be needed if other shift operators were used. 

Obviously,
$
{\mathrmd \over \mathrmdt} ( \sum_{n} \rho_n ) = 
   \sum_{n} {\dot{\rho}}_n = \sum_{n} (J_n - J_{n+1}),  
$ 
and this telescopic series vanishes if $J_n$ is bounded 
for all $n$ and $J_n$ vanishes at the boundaries. 
Then, $ \sum_{n} \rho_n $ is constant, and we have a quantity 
that is conserved in time.  

Let $D$ denote the {\it shift-down} operator and $U$ the 
{\it shift-up} operator. Both are defined on the set of all monomials.
If $m$ is a monomial then $D m = m |_{n \rightarrow n-1} $ 
and $U m = m |_{n \rightarrow n+1}. $ 
For example, $ D u_{n+2} v_{n} = u_{n+1} v_{n-1} $ and 
$ U u_{n-2} v_{n-1} = u_{n-1} v_{n}. $ 
It is easy to verify that compositions of $D$ and $U$ define an 
{\em equivalence relation\/} on monomials.
Simply stated, all shifted monomials are {\it equivalent}, e.g. 
$u_{n-1} v_{n+1} \, \equiv \, u_{n+2} v_{n+4} \, \equiv \, u_{n-3} v_{n-1}.$

In the algorithm below we will use the following 
{\it equivalence criterion}: if two monomials,
$m_1$ and $m_2$, are equivalent, $m_1 \, \equiv m_2,$ then
$m_1 \,=\, m_2 + [M_n - M_{n+1} ] $ for some polynomial $M_n$ 
that depends on ${\bf u}_n$ and its shifts.
For example, $ u_{n-2} u_n \, \equiv \, u_{n-1} u_{n+1} $
since $ u_{n-2} u_n = u_{n-1} u_{n+1} + [u_{n-2} u_{n} - u_{n-1} u_{n+1}] 
= u_{n-1} u_{n+1} + [M_n - M_{n+1}], $ with $M_n = u_{n-2} u_{n}.$

Also for later use, we call the {\it main} representative of an 
equivalence class, the monomial of that class with label $n$ on $u$ (or $v$). 
For example, $u_n u_{n+2}$ is the main representative of the class 
with elements $u_{n-1} u_{n+1}, u_{n+1} u_{n+3},$ etc. 
We use lexicographical ordering to resolve conflicts. For example, 
$u_n v_{n+2} $ (not $u_{n-2} v_n$) is the main representative in 
the class with elements $u_{n-3} v_{n-1}, u_{n+2} v_{n+4},$ etc.

\subsection{Algorithm}

To illustrate our algorithm, we consider the one-dimensional lattice 
\cite{MT,MH}
\begin{equation}\label{orgtoda}
{\ddot{y}}_n = \exp{(y_{n-1} - y_n)} - \exp{(y_n - y_{n+1})},
\end{equation}
due to Toda.
In (\ref{orgtoda}), $y_n$ is the displacement from equilibrium of the 
$n\/$th particle with unit mass under an exponential decaying interaction 
force between nearest neighbors. 

With the change of variables, 
\[ 
u_n = {\dot{y}}_n, \quad\quad\quad  v_n = \exp{(y_{n} - y_{n+1})}, 
\]
lattice (\ref{orgtoda}) can be written in algebraic form
\begin{equation}\label{todalatt}
{\dot{u}}_n = v_{n-1} - v_n, \quad\;\;\; {\dot{v}}_n = v_n (u_n - u_{n+1}).
\end{equation}
We can compute a couple of conservation laws for (\ref{todalatt}) by hand.
Indeed, 
$ \dot{u}_n = \dot{\rho}_n = v_{n-1} - v_n = J_n - J_{n+1} $ with
$ J_n = v_{n-1}. $
We denote this {\it first} pair by 
\[
\rho_n^{(1)} = u_n, \quad\quad\quad\; J_n^{(1)} = v_{n-1}. 
\]
After some work, we obtain a {\it second} pair:
\[
\rho_n^{(2)} = {\textstyle{1 \over 2}} {u_n}^2 + v_n, \quad\quad J_n^{(2)} 
= u_n v_{n-1}. 
\]

Key to our method is the observation that (\ref{todalatt}), and 
(\ref{conslaw}) together with the above densities and fluxes, 
are invariant under the scaling symmetry
\begin{equation}\label{symtoda}
(t, u_n, v_n) \rightarrow (\lambda t, \lambda^{-1} u_n, {\lambda}^{-2} v_n), 
\end{equation}
where $\lambda$ is an arbitrary parameter. The result of this
dimensional analysis can be stated as follows: 
$u_n$ corresponds to one derivative with respect to $t;$ for short, 
$ u_n \sim \frac{\mathrmd}{\mathrmdt}. $
Similarly, $v_n \sim \frac{{\mathrmd}^2}{{\mathrmdt}^2}. $
Scaling invariance, which is a special Lie-point symmetry, is an intrinsic 
property of many integrable nonlinear PDEs and DDEs.
Our algorithm exploits this property to find conserved densities, which now 
proceeds in three steps.
\vskip 5pt
\noindent
{\em Step 1: Determine the weights of variables}
\vskip 5pt
\noindent
The {\em weight\/}, $w,$ of a variable is by definition equal to the number 
of derivatives with respect to $t\/$ the variable carries. 
Weights are positive, rational, and independent of $n.$
We set $w({\mathrmd \over \mathrmdt}) = 1.$
In view of (\ref{symtoda}), we have $w(u_n) = 1,$ and $w(v_n) = 2.$ 

The {\em rank} of a monomial is defined as the total weight of the monomial, 
again in terms of derivatives with respect to $t$. 
For instance, the rank of each monomial in $\rho_n^{(2)}$ is two. 
Observe that in each equation of (\ref{todalatt}), all the terms (monomials) 
have the same rank. This property is called {\em uniformity in rank}. 
Densities and fluxes are also uniform in rank, and from (\ref{conslaw}), 
it follows that ${\rm rank}(J_n) = {\rm rank}(\rho_n) + 1, $ since 
$w(\frac{\mathrmd}{{\mathrmdt}}) = 1.$

Conversely, requiring uniformity in rank for each equation in
(\ref{todalatt}) allows one to compute the weights of the dependent
variables. Indeed,
$ w(u_n) + 1 \!=\! w(v_n), w(v_n) + 1 \!=\! w(u_n) + w(v_n), $ 
yields
$ w(u_n) = 1, \, w(v_n) = 2, $ which is consistent with (\ref{symtoda}).
% \vfill
% \newpage
\vskip 5pt
\noindent
{\em Step 2: Construct the form of the density}
\vskip 5pt
\noindent
As an example, let us compute the form of the density of rank $3.$
List all monomials in $u_n$ and $v_n$ of rank $3$ or less:
$ {\cal G} \!=\! \{ {u_n}^3, {u_n}^2, u_n v_n, {u_n}, v_n \}. $

Next, for each monomial in $\cal G$, introduce enough $t$-derivatives,
so that each term exactly has weight $3.$ Thus, using (\ref{todalatt}), 
\begin{eqnarray*}
&& 
\!\!\!\!\! 
{{\mathrmd}^0 \over {\mathrmdt}^0} ( {u_n}^3 )
   = {u_n}^3 , \;\;\;\;\quad 
{{\mathrmd}^0 \over {\mathrmdt}^0} ( u_n v_n )
   = u_n v_n , \\
&& \!\!\!\!\!  \\
&& \!\!\!\!\! 
{{\mathrmd} \over {\mathrmdt}} ( {u_n}^2 )
   = 2 u_n {\dot{u}}_n = 2 u_n v_{n-1} - 2 u_n v_n ,  \;\;\;\;\quad
{{\mathrmd} \over {\mathrmdt}} ( v_n )
   = {\dot{v}}_n =  u_n v_n -  u_{n+1} v_n, \\
&&  \!\!\!\!\!  \\
&& 
\!\!\!\!\! 
{{\mathrmd}^2 \over {\mathrmdt}^2} ( u_n ) 
   = {{\mathrmd} \over {\mathrmdt}} ( {\dot{u}}_n ) 
   = {{\mathrmd} \over {\mathrmdt}} ( v_{n-1} - v_n ) 
   = u_{n-1} v_{n-1} - u_{n} v_{n-1} - u_n v_n + u_{n+1} v_n .
\end{eqnarray*}
Gather the resulting terms in a set $\cal H=$
$
\{ {u_n}^3, u_n v_{n-1} , u_n v_n , u_{n-1} v_{n-1} , u_{n+1} v_n \} . 
$
Identify members that belong to the same equivalence classes and 
replace them by the main representatives. 
For example, since $ u_n v_{n-1} \equiv u_{n+1} v_n \,$ both are 
replaced by $u_n v_{n-1}.$ Doing so, $\cal H$ is replaced by 
$
{\cal I} =
\{ {u_n}^3 , u_n v_{n-1} , u_n v_n \}, 
$
which contains the building blocks of the density. 
Linear combination of the monomials in ${\cal I}$ with constant
coefficients $c_i$ gives the form of the density:
\begin{equation}\label{formrho3toda}
\rho_n = c_1 \, {u_n}^3 + c_2 \, u_n v_{n-1} + c_3 \, u_n v_n . 
\end{equation}
\vskip 3pt
\noindent
{\em Step 3: Determine the unknown coefficients in the density}
\vskip 5pt
\noindent
Now we determine the coefficients $c_1$ through $c_3$ by requiring that 
(\ref{conslaw}) holds. During this step we also compute the unknown 
flux $J_n.$ 

Compute ${\dot{\rho}}_n$ using (\ref{formrho3toda}). Then use 
(\ref{todalatt}) to remove ${\dot{u}_n}, {\dot{v}_n}, $ etc. 
After grouping the terms
\begin{eqnarray*}
{\dot \rho}_n 
& \,= \,& ( 3 c_1 - c_2 ) {u_n}^2 v_{n-1} + (c_3 - 3 c_1 ) {u_n}^2 v_n
          + (c_3 - c_2) v_{n-1} v_n  \\
& &       + c_2 u_{n-1} u_n v_{n-1} + c_2 {v_{n-1}}^2
          - c_3 u_n u_{n+1} v_n - c_3 {v_n}^2. 
\end{eqnarray*}
Use the equivalence criterion to modify ${\dot \rho}_n$.
For instance, replace $u_{n-1} u_n v_{n-1}$ by
$u_{n} u_{n+1} v_n + [u_{n-1} u_n v_{n-1} - u_{n} u_{n+1} v_{n}]$. 
The goal is to introduce the main representatives. 
Therefore,
\begin{eqnarray*}
{\dot \rho}_n 
& \,= \,& ( 3 c_1 - c_2 ) {u_n}^2 v_{n-1} + (c_3 - 3 c_1 ) {u_n}^2 v_n \\
& &+ (c_3 - c_2) v_{n} v_{n+1} + 
[(c_3 - c_2) v_{n-1} v_n -(c_3-c_2) v_{n} v_{n+1}] \\
& & + c_2 u_{n} u_{n+1} v_{n} + 
    [c_2 u_{n-1} u_n v_{n-1} - c_2 u_{n} u_{n+1} v_{n}]  \\
& & + c_2 {v_{n}}^2 + [c_2 {v_{n-1}}^2 - c_2 {v_{n}}^2 ] 
- c_3 u_n u_{n+1} v_n - c_3 {v_n}^2. 
\end{eqnarray*}
Next, group the terms outside of the square brackets and move the pairs
inside the square brackets to the bottom. Rearrange the latter terms 
so that they match the pattern $[J_n - J_{n+1}]$. Hence,
\begin{eqnarray*}
{\dot \rho}_n 
& \,= \,& ( 3 c_1 - c_2 ) {u_n}^2 v_{n-1} + (c_3 - 3 c_1 ) {u_n}^2 v_n \\
& & + (c_3 - c_2) v_{n} v_{n+1} + (c_2 - c_3) u_{n} u_{n+1} v_{n} 
+ ( c_2 - c_3 ) {v_n}^2 \\
& & + [ 
\{ 
(c_3 - c_2) v_{n-1} v_n + c_2 u_{n-1} u_n v_{n-1} + c_2 {v_{n-1}}^2  
\} \\
& & - \{ 
(c_3-c_2) v_{n} v_{n+1} + c_2 u_{n} u_{n+1} v_{n} + c_2 {v_{n}}^2 
\} ].
\end{eqnarray*}
The terms inside the square brackets determine:
\begin{equation} \label{formflux3toda}
J_n = (c_3 - c_2) v_{n-1} v_n + c_2 u_{n-1} u_n v_{n-1} + c_2 {v_{n-1}}^2. 
\end{equation}
The terms outside the square brackets must all vanish, yielding 
\begin{equation}\label{todasystem}
{\cal S} = \{ 3 c_1 - c_2 = 0, c_3 - 3 c_1 = 0, c_2 - c_3 = 0 \}. 
\end{equation}
The solution is $ 3 c_1 = c_2 = c_3.$ 
Since densities can only be determined up to a multiplicative constant, 
we choose $ c_1 = \frac{1}{3}, \, c_2 = c_3 = 1,$ 
and substitute this into (\ref{formrho3toda}) and (\ref{formflux3toda}). 
Hence,
\[
\rho_n = {\textstyle{1 \over 3}} \, {u_n}^3 + u_n ( v_{n-1} + v_n ),
\quad\;\;\; J_n = u_{n-1} u_n v_{n-1} + {v_{n-1}}^2.
\]
Analogously, we computed conserved densities of rank 
$\leq 5$ for (\ref{todalatt}). They are:
\begin{eqnarray*}
\rho_n^{(1)} &=& u_n, \quad\;\;
\rho_n^{(2)} = {\textstyle{1 \over 2}} {u_n}^2 + v_n, \quad\;\;
\rho_n^{(3)} = {\textstyle{1 \over 3}} {u_n}^3 + u_n (v_{n-1} + v_n), \\
&& \\
\rho_n^{(4)} &=& {\textstyle{1 \over 4}} {u_n}^4 + {u_n}^2 (v_{n-1} + v_n)
+ u_n u_{n+1} v_n + {\textstyle{1 \over 2}} {v_n}^2 + v_n v_{n+1}, \\
&& \\
\rho_n^{(5)} &=& {\textstyle{1 \over 5}} {u_n}^5 + {u_n}^3 (v_{n-1} + v_n ) 
+ u_n u_{n+1} v_n ( u_n + u_{n+1}) \\
&& + u_n v_{n-1} (v_{n-2} + v_{n-1} + v_n )
+ u_n  v_n ( v_{n-1} + v_n +  v_{n+1} ).
\end{eqnarray*}
Ignoring irrelevant shifts in $n,$ these densities agree with the 
results in \cite{MH}. 

To illustrate how our algorithm works for DDEs with parameters, consider
\begin{equation}\label{partodalatt}
{\dot{u}}_n = \alpha \; v_{n-1} - v_n, \quad 
{\dot{v}}_n = v_n \; (\beta \; u_n - u_{n+1}),
\end{equation} 
where $\alpha$ and $\beta$ are {\it nonzero} parameters.
In \cite{ARandBGandKT} it was shown that (\ref{partodalatt})
is completely integrable if $\alpha = \beta = 1.$

Using our algorithm, one can easily compute the {\it compatibility conditions}
for $\alpha$ and $\beta$, so that (\ref{partodalatt}) admits a polynomial 
conserved densities of, say, rank 3.
The steps are the same as for (\ref{todalatt}). 
However, (\ref{todasystem}) must be replaced by 
\[
\!\!\!\!\!\!\!\!\!\! 
{\cal S} = \{ 3 \alpha c_1 - c_2 = 0, \beta c_3 - 3 c_1 = 0, 
\alpha c_3 - c_2 = 0, \beta c_2 - c_3 = 0, \alpha c_2 - c_3 = 0 \}. 
\]
A non-trivial solution $ 3 c_1 = c_2 = c_3 $ will exist if and only if
$ \alpha = \beta = 1.$

Analogously, (\ref{partodalatt}) has density $\rho_n^{(1)} = u_n$ of rank 1 
if $ \alpha = 1, $ and density
$ \rho_n^{(2)} = \frac{\beta}{2} {u_n}^2 + v_n $ of rank 2 if 
$ \alpha \, \beta = 1.$ 
Only when $ \alpha = \beta = 1 $ will (\ref{partodalatt}) have
conserved densities of rank $\ge$ 3. 

\section{Examples}

\subsection{The Volterra Equation}

Consider the integrable discretization of the KdV equation:
\begin{equation} \label{kdv}
{\dot{u}}_n = u_n \, (u_{n+1}-u_{n-1}), 
\end{equation}
which is known as the Kac-Van Moerbeke equation or a special form of the 
Volterra system.
It arises in the study of Langmuir oscillations in plasmas, 
and in population dynamics \cite{MAandPC,MKandPM,MW}.

Notice that (\ref{kdv}) is invariant under the scaling symmetry
$(t, u_n) \rightarrow (\lambda t, {\lambda}^{-1} u_n).$
Hence, $u_n$ corresponds to one derivative with respect to $t,$ 
i.e. $u_n \sim \frac{\mathrmd}{\mathrmdt}.$
All terms in (\ref{kdv}) have the same rank if 
$ w({u_n}) + 1 = 2 \, w({u_n}),$ thus, $w({u_n}) = 1,$ 
which agrees with the scaling symmetry.

Let us find the form of density with rank $3.$
Forming all monomials of $u_n$ with rank $3$ or less yields the list
$ {\cal G} = \{ {u_n}^3 \, , \, {u_n}^2 , \, {u_n} \}. $
Introducing the necessary $t$-derivatives, leads to $\cal H$:
\[\!\!\!\!\!\!\!\!\!\!\!\!
\{ {u_n}^3,\! {u_n}^2 u_{n+1},\! u_{n-1} {u_n}^2 , \!
   u_n {u_{n+1}}^2, u_{n-1} u_n u_{n+1},\!
   u_n u_{n+1} u_{n+2},\! u_{n-2} u_{n-1} u_n,
   {u_{n-1}}^2 u_n\}. 
\]
Using $ u_{n-2} u_{n-1} u_n \equiv u_{n-1} u_n u_{n+1}$
$ \equiv u_n u_{n+1} u_{n+2},$ $ u_{n-1} {u_n}^2 \equiv u_n {u_{n+1}}^2 $
and $ {u_{n-1}}^2 u_n \equiv {u_n}^2 u_{n+1}, $ 
we obtain the list 
\[ 
{\cal I} =
\{ {u_n}^3 , {u_n}^2 u_{n+1} , u_n {u_{n+1}}^2 , u_n u_{n+1} u_{n+2} \}.
\]
A linear combination of the terms in $\cal I$ with constant
coefficients $c_i$ gives
\[ \label{formrho3}
\rho_n = c_1 \, {u_n}^3 + c_2 \, {u_n}^2 u_{n+1} + c_3 \, u_n {u_{n+1}}^2
+ c_4 \, u_n u_{n+1} u_{n+2}. 
\]
Proceed with step 3. After differentiation, shifting and regrouping
\begin{eqnarray} \label{volterra}
{\dot \rho}_n 
&=& (3 c_1 - c_2 ) {u_n}^3 u_{n+1} 
     + ( c_3 - 3 c_1 ) u_n {u_{n+1}}^3
     + 2 ( c_2 - c_3 ) {u_n}^2 {u_{n+1}}^2 \nonumber \\
& &  + 2 ( c_3 - c_2 ) u_n {u_{n+1}}^2 u_{n+2}
     + ( c_2 - c_4 ) {u_n}^2 u_{n+1} u_{n+2} \nonumber \\
& &  + ( c_4 - c_3 ) u_n u_{n+1} {u_{n+2}}^2 + [J_n - J_{n+1}],  
\end{eqnarray}
with 
\[
\!\!\!\!\!\!\!\!
J_n = - (3 c_1 u_{n-1} {u_n}^3 + 2 c_2 u_{n-1} {u_n}^2 u_{n+1}  
      + c_3 u_{n-1} {u_n} {u_{n+1}}^2 
      + c_4 u_{n-1} {u_n} u_{n+1} u_{n+2}). 
\]
The monomials outside the square brackets in (\ref{volterra}) must vanish. 
This yields 
\[
{\cal S} = \{ 3 c_1 - c_2 = 0, c_3 - 3 c_1 = 0, c_2 - c_3 = 0,
c_2 - c_4 = 0, c_4 - c_3 = 0 \}.
\]
Choosing $ c_1 = \frac{1}{3}, $ one has $ c_2 = c_3 = c_4 = 1. $
Therefore,
\begin{eqnarray*}
\rho_n &=& 
{\textstyle{1 \over 3}}\, {u_n}^3 + u_n u_{n+1} ( u_n + u_{n+1} + u_{n+2} ),\\
J_n &=& -( u_{n-1} {u_n}^3 + 2 u_{n-1} {u_n}^2 u_{n+1} 
      + u_{n-1} u_n {u_{n+1}}^2 + u_{n-1} u_n u_{n+1} u_{n+2} ).
\end{eqnarray*}
Analogously, for (\ref{kdv}) we computed the densities of rank $\leq 5:$ 
\begin{eqnarray*}
\!\!\!\! 
\rho_n^{(1)} &=& u_n, 
\quad\quad\quad\;\;\;\; 
\rho_n^{(2)} = {\textstyle{1 \over 2}} {u_n}^2 + u_n u_{n+1}, 
\\
\!\!\!\!
\rho_n^{(3)} &=& 
{\textstyle{1 \over 3}} {u_n}^3 + u_n u_{n+1} (u_n + u_{n+1} + u_{n+2} ),\\
\!\!\!\!
\rho_n^{(4)} &=& 
{\textstyle{1 \over 4}} {u_n}^4 + {u_n}^3 u_{n+1}
+ {\textstyle{3 \over 2}} {u_n}^2 {u_{n+1}}^2 
+ u_n {u_{n+1}}^2 (u_{n+1} + u_{n+2} ) \\
\!\!\!\!
& & + u_n u_{n+1} u_{n+2} ( u_n + u_{n+1} + u_{n+2} + u_{n+3} ), \\
\!\!\!\!
\rho_n^{(5)} &=& 
{\textstyle{1 \over 5}} {u_n}^5 + u_n u_{n+1} ( {u_n}^3 + {u_{n+1}}^3 )
+ 2 {u_n}^2 {u_{n+1}}^2 ( u_n + u_{n+1} ) \\
&& + u_n u_{n+1} u_{n+2} ({u_n}^2 + u_n u_{n+2} + u_{n+1} u_{n+3} )
+ 3 u_n {u_{n+1}}^2 u_{n+2} \\ 
&& ( u_n + u_{n+1} + u_{n+2}) + u_n u_{n+1} {u_{n+2}}^2 (u_{n+2} + u_{n+3}) \\
&& + u_n u_{n+1} u_{n+2} u_{n+3} (u_n + u_{n+1} + u_{n+2} + u_{n+3} + u_{n+4}).
\end{eqnarray*}
\subsection{Discretizations of the nonlinear Schr\"odinger equation}

In \cite{MAandJLa,MAandJLb}, Ablowitz and Ladik studied properties of the 
following integrable discretization of the NLS equation:
\begin{equation} \label{orgabllad}
i \, {\dot{u}}_n =
u_{n+1}- 2 u_n + u_{n-1} \pm  u_n^{*} u_n (u_{n+1} + u_{n-1}), 
\end{equation}
where $u_n^{*}$ is the complex conjugate of $u_n.$
We continue with the $+\;$ sign; the other case is analogous.
Instead of splitting $u_n$ into its real and imaginary parts, we treat
$u_n$ and $v_n = u_n^{*}$ as independent variables and 
augment (\ref{orgabllad}) with its complex conjugate equation. 
Absorbing $i$ in the scale on $t,$ we get
\begin{eqnarray} \label{abllad}
{\dot{u}}_n 
&=& u_{n+1} - 2 u_n + u_{n-1} + u_n v_n (u_{n+1} + u_{n-1}), 
\nonumber \\ 
{\dot{v}}_n 
&=& -( v_{n+1} - 2 v_n + v_{n-1} ) -  u_n v_n (v_{n+1} + v_{n-1}).
\end{eqnarray}
Since $v_n = u_n^{*}, $ we have $w(v_n) = w(u_n).$ 

Neither of the equations in (\ref{abllad}) is uniform in rank. 
To circumvent this problem we introduce an auxiliary parameter 
$\alpha$ with weight, and replace (\ref{abllad}) by
\begin{eqnarray} \label{ablladnew}
{\dot{u}}_n 
&=& \alpha ( u_{n+1} - 2 u_n + u_{n-1} ) + u_n v_n (u_{n+1} + u_{n-1}), 
\nonumber \\
{\dot{v}}_n 
&=& - \alpha ( v_{n+1} - 2 v_n + v_{n-1} ) -  u_n v_n (v_{n+1} + v_{n-1}).
\end{eqnarray}
Uniformity in rank requires that
\begin{eqnarray*}
w(u_n) + 1 &=& w(\alpha) + w(u_n) = 2 w(u_n) + w(v_n) = 3 w(u_n), \\
w(v_n) + 1 &=& w(\alpha) + w(v_n) = 2 w(v_n) + w(u_n) = 3 w(v_n), 
\end{eqnarray*}
which yields $ w(u_n) = w(v_n) = \frac{1}{2}, w(\alpha) = 1, $
or, $ {u_n}^2 \sim {v_n}^2 \sim \alpha \sim {\mathrmd \over \mathrmdt}.$

Recall that the uniformity in rank requirement is essential for the first
two steps of the algorithm. However, after Step 2, we may set $\alpha = 1.$
The computations now proceed as in the previous examples. 
We list some conserved densities of (\ref{abllad}), 
which correspond with those in \cite{MAandJLa}:
\begin{eqnarray*}
\!\!\!\!\!\!
\rho_n^{(1)} &=& c_1 u_n v_{n-1} + c_2 u_n v_{n+1}, \\ 
\!\!\!\!\!\!
\rho_n^{(2)} &=& c_1 
(
{\textstyle{1 \over 2}} {u_n}^2 {v_{n-1}}^2 + u_n u_{n+1} v_{n-1} v_n + u_n v_{n-2} 
) \\
\!\!\!\!\!\!
&+& c_2 (
{\textstyle{1 \over 2}} {u_n}^2 {v_{n+1}}^2 + u_n u_{n+1} v_{n+1} v_{n+2} + u_n v_{n+2}
), \\
\!\!\!\!\!\!
\rho_n^{(3)} &=& c_1 [
{\textstyle{1 \over 3}} {u_n}^3 {v_{n-1}}^3 + 
u_n u_{n+1} v_{n-1} v_n ( u_n v_{n-1} + u_{n+1} v_n + u_{n+2} v_{n+1} ) \\
\!\!\!\!\!\!
&+& u_n v_{n-1} (u_n v_{n-2} + u_{n+1} v_{n-1} )
+ u_n v_n (u_{n+1} v_{n-2} + u_{n+2} v_{n-1} ) + u_n v_{n-3} ] \\
\!\!\!\!\!\!
&+&  c_2 [ 
{\textstyle{1 \over 3}} {u_n}^3 {v_{n+1}}^3 + 
u_n u_{n+1} v_{n+1} v_{n+2} (u_n v_{n+1} +u_{n+1} v_{n+2} + u_{n+2} v_{n+3}) \\
\!\!\!\!\!\!
&+& u_n v_{n+2} (u_n v_{n+1} + u_{n+1} v_{n+2} )
+ u_n v_{n+3} (u_{n+1} v_{n+1} + u_{n+2} v_{n+2} ) + u_n v_{n+3}].
\end{eqnarray*}
As shown in \cite{MAandJLa}, scheme (\ref{orgabllad}), if defined on 
an infinite interval, admits infinitely many independent conserved densities.
Although it is a constant of motion, we cannot derive the Hamiltonian of 
(\ref{orgabllad}), for it has a logarithmic term \cite{MAandBH}. 

We also computed conserved densities of the non-integrable {\it standard}
second-order scheme \cite{MAandBH}, 
\begin{equation} \label{orgstandnls}
i \, {\dot{u}}_n =
u_{n+1} - 2 u_n + u_{n-1} + 2 u_n^{*} u_n^2,
\end{equation}
for the NLS equation. 
Instead of (\ref{ablladnew}) one has
\begin{eqnarray*}
{\dot{u}}_n 
&=& \alpha (u_{n+1} - 2 u_n + u_{n-1}) + 2 u_n^2 v_n, \\
{\dot{v}}_n 
&=& - \alpha (v_{n+1} - 2 v_n + v_{n-1}) - 2 u_n v_n^2. 
\end{eqnarray*}
Here,  
$
{u_n}^2 \sim {v_n}^2 \sim \alpha \sim {\mathrmd \over \mathrmdt}. $
We could only find two independent conserved densities. 
Indeed, after setting $\alpha = 1,$
\[ 
\rho_n^{(1)} = u_n v_n,  \quad\quad\quad\;\;\;
\rho_n^{(2)} = {u_n}^2 {v_n}^2 + u_n v_{n-1} + u_n v_{n+1}. 
\]
% This indicates that (\ref{orgstandnls}) is non-integrable 
% \cite{MAandBH,MAandJLa,MAandJLb}. 

\subsection{Generalized Toda lattices} 

Recently, Suris \cite{BD,YBSa} showed the integrability of the chain
\begin{equation}\label{reltoda}
{\ddot{y}}_n = 
\dot{y}_{n+1} e^{(y_{n+1} - y_n)} - e^{2(y_{n+1} - y_n)} 
- \dot{y}_{n-1} e^{(y_{n} - y_{n-1})} + e^{2 (y_{n} - y_{n-1})},
\end{equation}
which is related to the relativistic Toda lattice. 
With the change of variables, 
$u_n = {\dot{y}}_n, \;  v_n = \exp{(y_{n+1} - y_n)},$
(\ref{reltoda}) can be written as
\begin{equation} \label{reltodalatt}
{\dot{u}}_n = 
v_n (u_{n+1} - v_n) - v_{n-1} (u_{n-1} - v_{n-1}), \quad\;\;
{\dot{v}}_n = v_n (u_{n+1} - u_n).
\end{equation}
Here, $ u_n \sim v_n \sim \frac{\mathrmd}{\mathrmdt}, $ 
and we computed five conserved densities for (\ref{reltodalatt}).
The first three are:
\begin{eqnarray*}
\rho_n^{(1)} &=& u_n - v_n, \quad\quad\quad\quad
\rho_n^{(2)} = {u_n}^2 - {v_n}^2,  \\
\rho_n^{(3)} &=& {\textstyle{1 \over 3}} ( {u_n}^3 + 2 {v_n}^3 )
- u_n ( {v_{n-1}}^2 + {v_n}^2 ) + u_n u_{n+1} v_n.
\end{eqnarray*}
Suris \cite{YBSb} also investigated
\begin{equation}\label{backtoda}
{\ddot{y}}_n = 
\dot{y}_n \left[\exp{(y_{n+1} - y_n)} - \exp{(y_{n} - y_{n-1})} \right],
\end{equation}
which is closely related to the classical Toda lattice (\ref{orgtoda}).
The same change of variables as for (\ref{reltoda}) allows one to
write (\ref{backtoda}) as
\begin{equation} \label{backtodalatt}
{\dot{u}}_n = u_n (v_n - v_{n-1}), \quad\quad
{\dot{v}}_n = v_n (u_{n+1} - u_n).
\end{equation}
Again, $ u_n \sim v_n \sim \frac{\mathrmd}{\mathrmdt}, $ and 
the first four conserved densities are
\begin{eqnarray*}
\!\!\!\!\!\!\!\!\!\!\!\!\!\!\!\!
\rho_n^{(1)} &=& u_n + v_n, \quad\quad\quad\quad
\rho_n^{(2)} = 
{\textstyle{1 \over 2}} ({u_n}^2 + {v_n}^2) + u_n (v_{n-1} + v_{n}), \\
\!\!\!\!\!\!\!\!\!\!\!\!\!\!\!\!
\rho_n^{(3)} &=& 
{\textstyle{1 \over 3}} ( {u_n}^3 + {v_n}^3 ) 
+ {u_n}^2 ( v_{n-1} + v_n ) + u_n ( {v_{n-1}}^2  +  {v_n}^2) 
+ u_n v_n (v_{n-1} + u_{n+1}), \\
\!\!\!\!\!\!\!\!\!\!\!\!\!\!\!\!
\rho_n^{(4)} &=& {\textstyle{1 \over 4}} ( {u_n}^4 + {v_n}^4 ) 
+ {u_n}^3 ( v_{n-1} + v_n ) + u_n ( {v_{n-1}}^3 + {v_n}^3 )
+ {\textstyle{3 \over 2}} {u_n}^2 ( {v_{n-1}}^2 + {v_n}^2 )  \\
&& + {u_n} u_{n+1} v_n (u_n + u_{n+1} ) + 
2 u_n v_n ( u_n v_{n-1}  + u_{n+1} v_n ) \\
&&+ u_n v_{n-1} v_n ( v_{n-1} + v_n ) + u_n u_{n+1} v_n ( v_{n-1} + v_{n+1}).
\end{eqnarray*}
\section{Conclusions}

We developed a {\it Mathematica} program, called {\it diffdens.m}, 
and used it to compute all the conserved densities presented in this paper. 
For lattices with parameters, the code automatically determines the 
compatibility conditions on these parameters so that a sequence of 
conserved densities might exist.

The existence of a large number of conservation laws is an indicator of 
integrability of the system.
Therefore, by generating the compatibility conditions, one can analyze 
classes of parameterized DDEs and filter out the candidates for complete 
integrability.

Future generalizations of the algorithm will exploit other symmetries in
the hope to find conserved densities of non-polynomial form. 

% \vfill
% \newpage
\vskip 5pt
\noindent
{\bf Acknowledgements}
\vskip 5pt
\noindent
We would like to acknowledge a helpful discussion with Dr. S. Chakravarty, 
who pointed us to relevant literature. 
We also thank Prof. F. Verheest for his insightful comments.


\begin{thebibliography}{30}
\bibitem{UGa} 
\"{U}. G\"{o}kta\c{s} and W. Hereman, The paper gokthere.ps is available 
via anonymous ftp from mines.edu in 
directory pub/papers/math\_cs\_dept/software/condens.
\end{thebibliography}
\end{document}